\documentclass[preprint,12pt,a4paper]{elsarticle}
\usepackage{amssymb,natbib,xcolor,url}

\usepackage{siunitx}
\usepackage{amsmath,amssymb,amsfonts}
\usepackage{graphicx}

\begin{document}

\begin{frontmatter}
	
\title{Charge capacity characteristics of a Lithium Nickel-Cobalt-Aluminium Oxide battery show fractional-derivative behavior}

\author[1]{Marcus T. Wilson$^*$} 
\author[2]{Vance Farrow}
\author[2]{Caleb Pyne}
\author[2]{Jonathan Scott}

\address[1]{Te Aka M\~atuatua -- School of Science, The University of Waikato, Hamilton, New Zealand}
\address[2]{School of Engineering, The University of Waikato,  Hamilton, New Zealand}
\address{*Corresponding author: Te Aka M\={a}tuatua--School of Science,
	University of Waikato,
	Private Bag 3105, Hamilton 3240, New Zealand; telephone +6478384834;
	fax +6478384835;
	email marcus.wilson@waikato.ac.nz}

\date{30 September 2021}

\begin{abstract}
Batteries experience capacity offset where available charge depends on the rate at which this charge is drawn. 
In this work we analyze the capacity offset of a \SI{4.8}{\ampere\hour} lithium nickel-cobalt-aluminium oxide battery using an equivalent circuit model of a fractional capacitor
in series with a resistor.  
In this case, the available charge, in theory, becomes infinite in the limit of infinitesimal rate. We show that the
fractional properties of the capacitor can be extracted from the charge against rate plot. We then use a network of RC elements
to represent the fractional capacitor in order to simulate the data with Matlab. We find that the fractional
exponent $\alpha$ obtained in this way, 0.971, agrees with that obtained in a more traditional manner from an impedance versus frequency plot, although the fractional capacity does not. Such an approach demonstrates the importance of a fractional
description for capacity offset even when an element is nearly a pure capacitor and is valuable
for predictions of state-of-charge when low currents are drawn.

\end{abstract}


\end{frontmatter}

\section{Introduction}
\label{se:introduction}

It has long been acknowledged that rechargeable batteries exhibit fractional-derivative characteristics; 
Randles showed this in 1947 by showing a good fit to data with his equivalent-circuit impedance model that included a Warburg element~\cite{randles_kinetics_1947}.
Sabatier et al.\ have shown good results in practical applications~\cite{sabatier_2006,cugnet_2010,sabatier_2013}.
More recently Westerhoff~\cite{westerhoff_analysis_2016} presented a review of battery models 
leading to an equivalent-circuit model (ECM) with five fractional-derivative branches. 

A fractional capacitor, frequently called a constant-phase element or CPE,
is a component whose terminal current and voltage are related by a fractional-power derivative (between zero and one). 
A Warburg element, common in battery ECMs, is a CPE of order exactly one half. This is usually attributed to 
diffusion of ions within the battery. 
It is distinct from the CPE that is responsible for the bulk of the energy store in a battery. 
These components can be described mathematically through fractional calculus~\cite{Samko_FractionalDiffintegrals}.
In the case of lithium-ion batteries incorporating cobalt, measurements of impedance suggest that
the order of the derivative is close to or greater than 0.8. 
This is in contrast with a value of approximately 0.5 measured for lead-acid and 0.7 for small lithium-iron-phosphate
batteries~\cite{scott_new_2019,hasan_fractional_2016,JianbangME}.

It is well known that a battery exhibits `capacity offset'. Here, the charge available to be drawn from a fully-charged battery depends on the rate at which it is drawn; faster rates (that is, higher discharge currents) result in a lower capacity. 
The phenomenon is commonly associated with Peukert's empirical equation~\cite{Doerffel_Peukert_review}.
While capacity offset has been explored at high currents, its nature at very low currents has not been demonstrated. A bizarre mathematical implication of fractional calculus is that a CPE can store, in principle, infinite energy at a finite voltage. This apparent paradox is resolved by understanding that a CPE needs infinite time to store this energy. The  behavior of a CPE for long-time processes in a battery (i.e. charge and discharge at very low currents) is therefore expected to be non-trivial. Specifically, we expect a considerable rise in capacity at very low currents.  

In this manuscript, we have for the first time measured the capacity offset and impedance of a \SI{4.8}{\ampere\hour}, 21650-format, 
lithium nickel-cobalt-aluminium oxide (NCA) battery, commonly used in traction applications, at ultra-long time scales. Specifically, we have taken 
impedance data to frequencies as low as \SI{0.5}{\micro\hertz} (i.e. a period of $2 \times 10^6$ seconds, approximately 23 days), 
and performed constant current charge-discharge cycles with periods as low as 8 days. We demonstrate that
the capacity offset behavior at low currents (long times) is consistent with predictions of fractional calculus, and
provides evidence of a CPE with power close to 1.

\section{Fractional Properties}

\subsection{Theory of fractional elements}

A fractional capacitor, or constant phase element (CPE) exhibits an admittance spectrum with a fractional power
$\alpha$, $0 < \alpha < 1$, and a constant phase between the voltage and current of $-\alpha \pi/2$ radians or 
$-\alpha \times 90^{\circ}$. 
The special case of $\alpha = 1$ is a pure capacitor, while $\alpha = 0$ is a resistor. 
Their behavior can be defined through fractional calculus. Generally, the current and 
voltage follow the relationship:
\begin{equation}
I(t) = C_F \frac{d^\alpha V(t)}{dt^\alpha},
\label{eq:CFdef}
\end{equation}
where the right-hand-side involves a \emph{fractional derivative}.
In this work it is more useful to consider how voltage $V(t)$ depends on current, and
we can invert this relationship 
 through a Riemann-Liouville integral, specifically (for $0 < \alpha < 1$):
\begin{equation}
V(t)=\frac{1}{C_F \Gamma(\alpha)}\int_{t'=\tau}^{t} I(t') (t-t')^{\alpha-1} dt' 
\label{eq:RL}
\end{equation}
where $\Gamma(\alpha)$ is the gamma-function and $\tau$ is a starting point in the past. 
One notable feature is that the fractional-derivative is not local --- the element has memory of its past~\cite{Westerlund_capacitor_1994}.

The  impedance of a fractional capacitor expressed as a function of frequency in the Laplace form follows naturally from Eq.~(\ref{eq:CFdef}):
\begin{equation}
Z_{CPE} = \frac{1}{C_F (j \omega)^\alpha}
\label{eq:Zdef}
\end{equation}
where $j^2=-1$.

\subsection{Impedance measurements on NCA \SI{4.8}{\ampere\hour} battery}

\begin{figure}[tbp]
	\centering
	\includegraphics[width=0.95\linewidth]{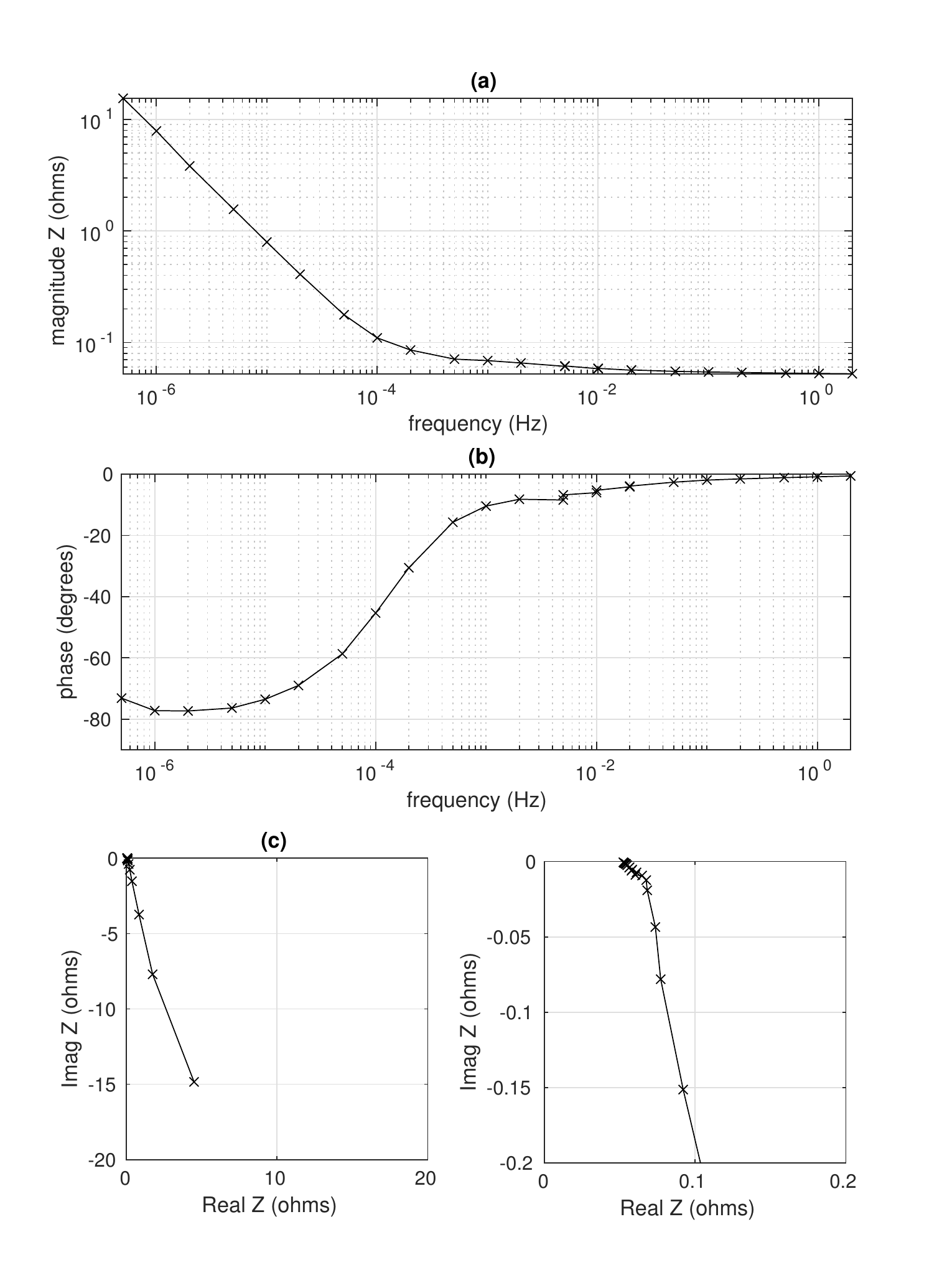}
	\caption{Impedance for a \SI{4.8}{\ampere\hour} NCA battery. (a) The magnitude of the
	impedance against frequency. (b) The phase of the impedance against frequency. 
(c) $\Re{Z}$ against $\Im{Z}$. The right-hand panel is a close-up of the high-frequency (low impedance) region. }
	\label{fig:impedancef}
\end{figure}

Figure~\ref{fig:impedancef} shows the impedance of a NCA \SI{4.8}{\ampere\hour} battery across a frequency range \SI{0.5}{\micro \hertz} to \SI{2}{\hertz}.
These data were obtained using a two-quadrant 66332A power supply measuring voltage and current over three cycles at $\approx $\SI{100}{\milli \A}
using the method described in~\cite{scott_new_2019} by means of software developed previously by one of us~\cite{farrow_characterisation_2020}. 
The plot covers ultra-low frequencies; the lowest frequency corresponds to a period of \num{2e6}~s, approximately 23 days. 
Part (a) shows the magnitude of impedance $Z$ against frequency, on a log-log scale, part (b) gives the phase against frequency, and part (c) 
shows the imaginary part of the impedance plotted against its real part. A CPE element is evidenced by the straight line with
gradient close to (but not equal to) $-1$ on the magnitude plot (a), a phase approaching approximately $-80^\circ$
at low frequency on the phase plot (b), and an approximately straight line (not quite vertical)
on part (c). 
Note that the phase of the lowest frequency measurement begins to curl back towards zero. 
This hints at the presence of a high-value, parallel resistance. The value is sufficiently high that we can ignore it in this work. 
 
It is also evident that at high frequency the impedance approaches a constant value, and the phase approaches zero, 
indicative of a resistor. This data suggests a model for the battery consisting of a fractional element in series with
a resistor. This, however, is not quite sufficient to describe the impedance spectrum. We note from (a) that the `knee' of the 
curve is much broader than expected for such a circuit. Moreover, we can see in the upper-part of the right-hand panel of part (c) evidence
of another constant phase element, this one having a constant phase of about $-45^\circ$ (the angle between the locus 
and the horizontal axis is about $-45^{\circ}$). This is likely to be the Warburg nature of the battery appearing. Thus a better model for the impedance data is a fractional element in series with a Warburg element and a resistor.
We emphasize however that it is likely that various different combinations of these three elements could also fit the 
data well~\cite{berthier_distinguishability_2001,EdenME}; indeed a vast number of different equivalent circuits have been proposed for batteries in the
literature. 
 
Fitting a straight line to the lowest 7 frequency points in part (a) yields parameters $\alpha=0.976(8)$ and $C_F=1.54(6) \times 10^{4}$~A~s$^{\alpha}$~V$^{-1}$,
where the number in parentheses is the uncertainty in the final digit. Fitting an asymptote to the high-frequency end of the
spectrum gives $R_s = 57$~m$\Omega$.

\subsection{Capacity of battery versus discharge rate}

In what follows we shall ignore the Warburg nature of the battery and concentrate on the simpler (if not complete) 
model of a CPE in series with a resistor, which we denote by CPE-R.

Traditionally  a  capacity vs C-rate  (i.e. current) graph is plotted as a log-scale for C-rate to space out the data points 
but this obscures useful information. 
It is straightforward to show that the capacity of a battery (that is, the amount
of charge that can be drawn for the voltage to fall from a defined high point of $V_h$ to a defined
low point of $V_l$)  against C-rate (current drawn) leads to a straight-line plot on a \emph{linear} scale, 
for the circuit model
of a \emph{pure} capacitor in series with a resistor.  The intercept on the $y$-axis is 
the capacity.

For the case of a \emph{fractional} capacitor, the behavior is more complicated. A fractional element can hold in principle
infinite charge, if it is charged and discharged infinitesimally slowly. Thus a CPE-R model leads to a charge versus current
plot that is straight at higher currents but bends upwards at the lowest currents. In what follows we present a
mathematical analysis of this effect using the Riemann-Liouville integral (\ref{eq:RL}).

 In principle, the behavior of the battery depends on its entire history. 
 This is intractable. Instead, we will think of the simplified case of charging the battery
  with current $I_0$ from time 0 to time $T$ then drawing out $-I_0$ for time $T$. 
  That is, Eq.~(\ref{eq:RL}) for the voltage $V_c$ over the fractional capacitor becomes, for $T \leq t \leq 2T$:
  \begin{equation}
  V_c(t) = \frac{1}{C_F \Gamma(\alpha)} \left[ \int_{t'=0}^{T} I_0 (t-t')^{\alpha-1} dt' 
 -\int_{t'=T}^{t} I_0 (t-t')^{\alpha-1} dt' \right]
   \label{eq:DeltaVcalc}
  \end{equation}
  which is
  \begin{equation}
  V_c(t)=\frac{I_0}{C_F \Gamma(\alpha+1)} \left[ t^\alpha - 2(t-T)^\alpha \right].
  \end{equation}
  
 Identifying the total voltage $V_h$ over the battery 
 at the end of the `charge' cycle as $V_h = V_c(T) + I_0 R_s$  where we have now included the voltage over the series resistor $R_s$, 
  and the voltage $V_l$ at the end of the `discharge' cycle as $V_l = V_c(2T) - I_0 R_s$, we can find the total voltage change on the discharge as
  \begin{equation}
  \Delta V = V_h - V_l  = \frac{(3-2^\alpha) I_0 T^\alpha}{C_F \Gamma(\alpha+1) } +2 I_0 R_s. 
  \end{equation} 
  We can thus find the discharge time $T$ as a function of the voltage change, 
  $\Delta V$, and then the charge moved as simply $Q = I_0 T$. This yields:
 \begin{equation}
 Q = \left[ \frac{C_F \Gamma(\alpha+1)}{3-2^\alpha}\left(\Delta V - 2 I_0 R_s\right)  \right]^{1/\alpha} I_0^{1-\frac{1}{\alpha}} 
 \label{eq:Qcalc}
 \end{equation}
 and so we expect in the limit of $I_0 \rightarrow 0$ that $Q \propto I_0^{1 - \frac{1}{\alpha} }$. 
 This is equivalent to Peukert's empirical equation, $T I_0^n={\rm const}$, with $n=1/\alpha$,
 although the literature usually discusses Peukert's relationship in terms of higher currents~\cite{Doerffel_Peukert_review}.  
 Thus a plot of $\log Q$ versus $\log I_0$ will have gradient
 $1-1/\alpha$ in the limit of $I_0 \rightarrow 0$ allowing identification of $\alpha$.
 $C_F$ can also be determined from the intercept with the y-axis. This gives us a method of measuring 
 the fractional element that complements that of the impedance spectrum (Fig.~\ref{fig:impedancef}).
 
 \label{sec:theory}

 \subsection{Modeling a fractional element}
 
 An unpleasant feature of the Riemann-Liouville integral~(\ref{eq:RL}) is that the 
 behavior of the voltage at time $t$ does not depend only on the current in the
 circuit at time $t$, but rather the whole history of the current. This means that
 simply iterating the voltage forward in time under a particular current-time profile,
 as one can do for a pure capacitor,
 is not possible. 
However, a fractional capacitor can be modeled in terms of an infinite array of R-C elements in parallel, 
as described in Morrison~\cite{morrison_rc_1959}, applied in~\cite{scott_compact_2013}, with corrections noted in~\cite{seshadri_correction_2018}. 
The concept here is that each branch has an RC-time constant, and these time constants
are logarithmically distributed across the branches. The voltage over this network
can be  incrementally solved against time for a given current input. 
Details of the network used in this modeling work are shown in the appendix.

 \section{Experimental method}
 
 \subsection{Measuring capacity against discharge rate}
 
The battery was cycled 
between 4.30~V and 3.00~V 
on a constant current cycle (equal and opposite charge and discharge currents), using the 66332A to provide
the controlled current and logging of voltage against time. 
Currents of 
\SI{5}{\A}, 2~A, 1~A, 0.5~A, 0.2~A, 0.1~A and 0.05~A 
were used, in this order. 
The capacity of the battery was identified as the charge drawn in the final discharge.  
The values for the circuit elements  $\alpha$ and $C_F$ for the fractional capacitor, and $R_s$ the series resistance, were extracted 
as described in Sec.~\ref{sec:theory} above. 

 \subsection{Modeling the battery with Matlab}
 
 After extracting values $\alpha$ and $C_F$ and $R_s$ from
 the capacity against current plot, we modeled the response of the battery under a charge-discharge cycle using Matlab. 
 We have used the Morrison description of the CPE, a parallel network of RC elements, as detailed in the appendix, and solved for the voltage over the CPE against time. The model results were compared with the experimental measurements.

 \section{Results}
 
\begin{figure}[tbp]
 	\centering
 	\includegraphics[width=0.95\linewidth]{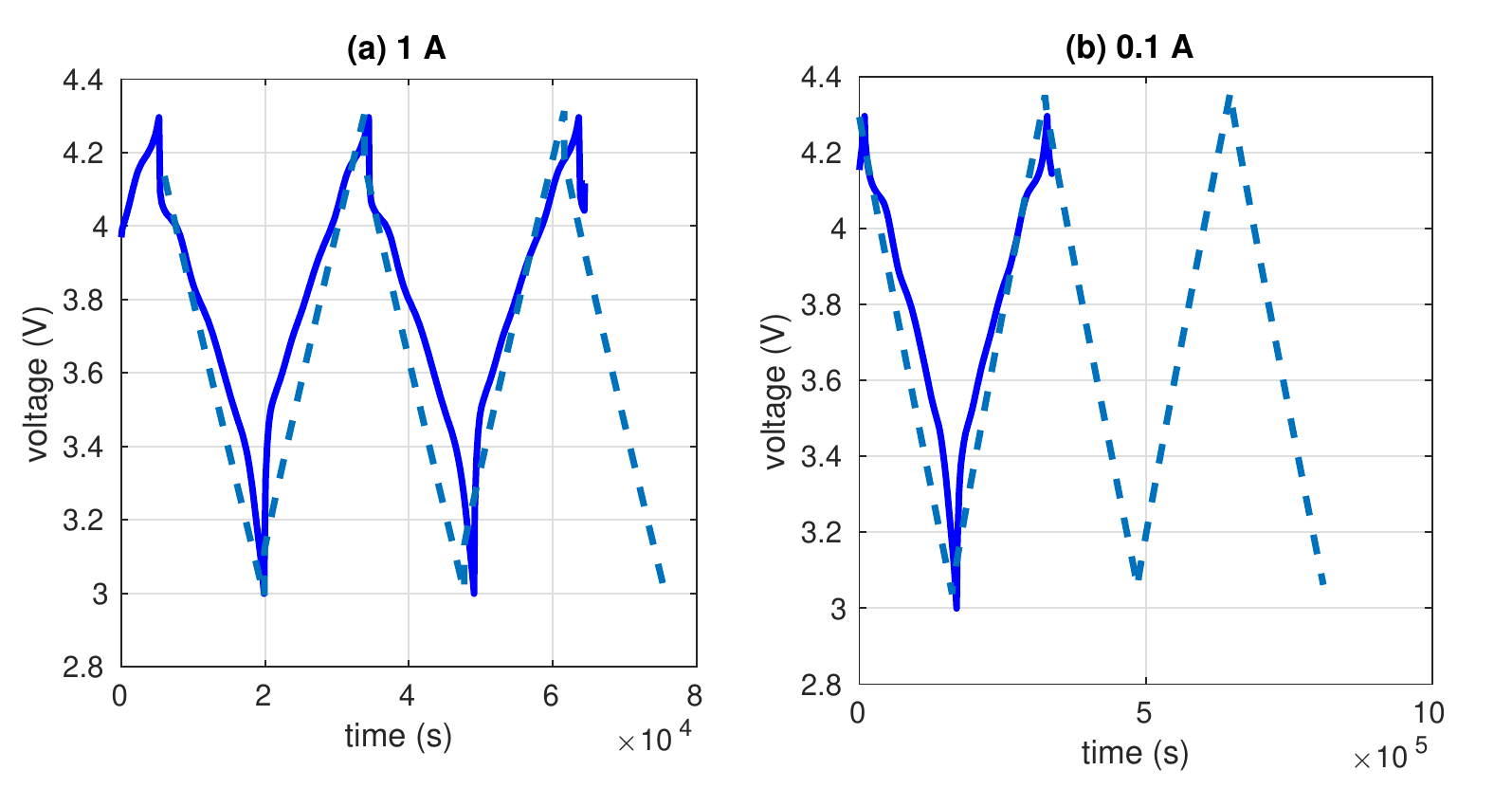}
 	\caption{A plot of the voltage on the battery against time, for a constant
 		discharge and charge current of (a) 1.0~A and (b) 0.1~A. The solid line is the experimental
 		result; the dashed line is the simulated data with $\alpha$ and $C_F$
 		extracted from the plot of Fig.~\ref{fig:Q_vs_I}.}
 	\label{fig:V_vs_t}
 \end{figure}

 In Fig.~\ref{fig:V_vs_t} we plot with the solid line the measured
 voltage against time trace for the final  
 discharge-charge cycle, for example currents of (a) 1.0~A and (b) 0.1~A. 
 The charge capacity is simply the integral of current over the discharge period, in 
 other words the time for discharge multiplied by the current used. 

\begin{figure}[tbp]
 	\centering
 	\includegraphics[width=0.95\linewidth]{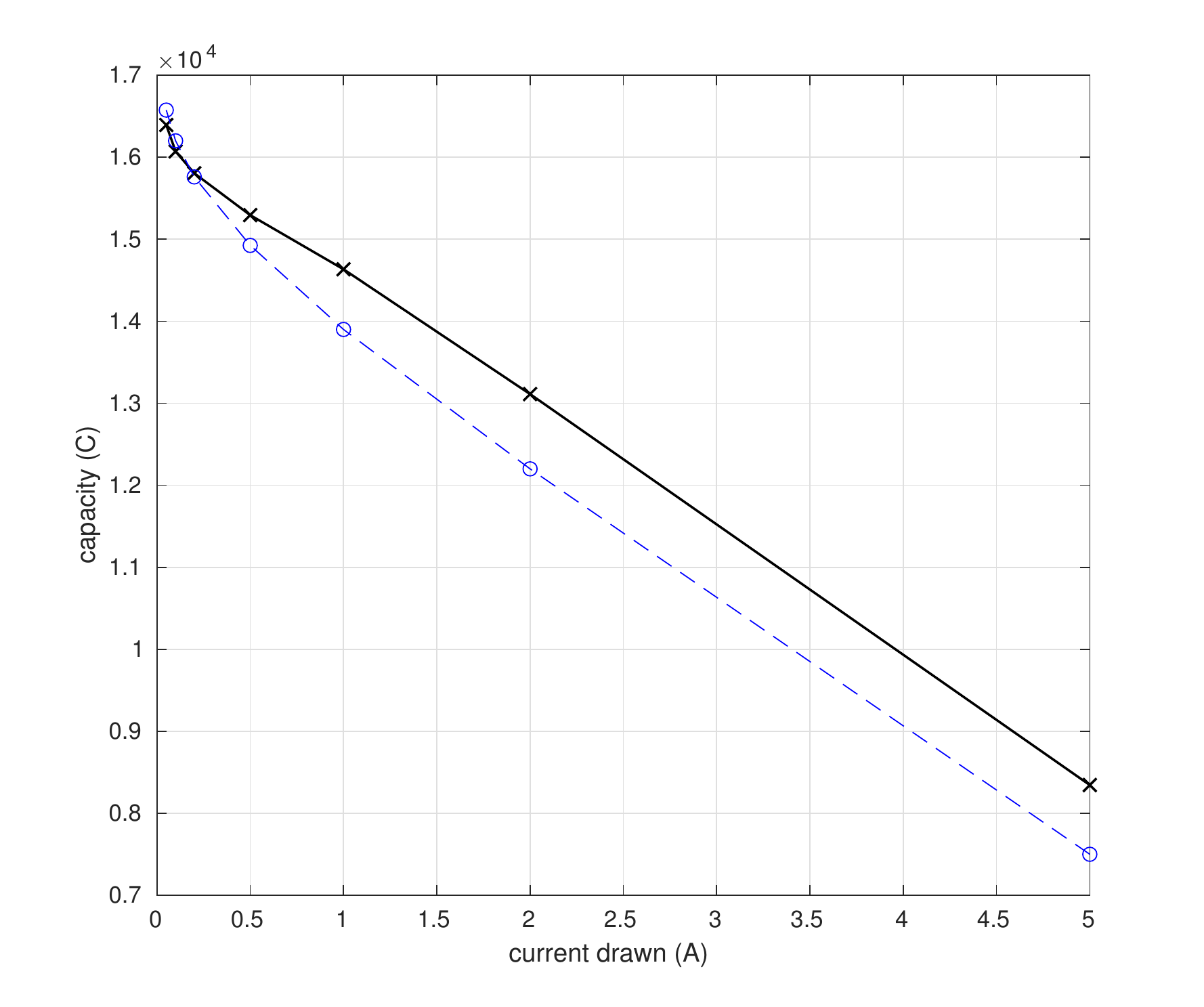}
 	\caption{A linear-scale plot of the capacity of the battery as a function
 		of the discharge current. The black solid line shows the experimentally-measured values; the
 	blue dashed the modeled results.}
 	\label{fig:Q_vs_I}
 \end{figure}

 The capacity has been extracted from these plots (and others like them for the other currents)
 and plotted as a function of current in  Fig.~\ref{fig:Q_vs_I}.
 The fractional properties $\alpha$ and $C_F$ are then extracted from this plot via
 Eq.~(\ref{eq:Qcalc}). By fitting a straight line to the data at the lowest four currents,
 we obtain $\alpha=0.9711(17)$, 
  and $C_F = 9.20(13)\times 10^3$~A~s$^{\alpha}$~V$^{-1}$.
  The series resistance $R_s$ is found from the $x$-intercept $I_x$ of the linear-scale plot of 
  Fig.~\ref{fig:Q_vs_I}; Eq.~(\ref{eq:Qcalc}) implies that capacity drops to zero ($x$-intercept) 
  when $I_0 = \Delta V / 2 R_s$, and so $R_s = \Delta V/(2 I_x)$ where $\Delta V$ for these
  measurements is 4.30~V $-$ 3.00~V = 1.3~V. Thus  $R_s = 0.0631$~$\Omega$.

 \begin{figure}[tbp]
 	\centering
 	\includegraphics[width=0.95\linewidth]{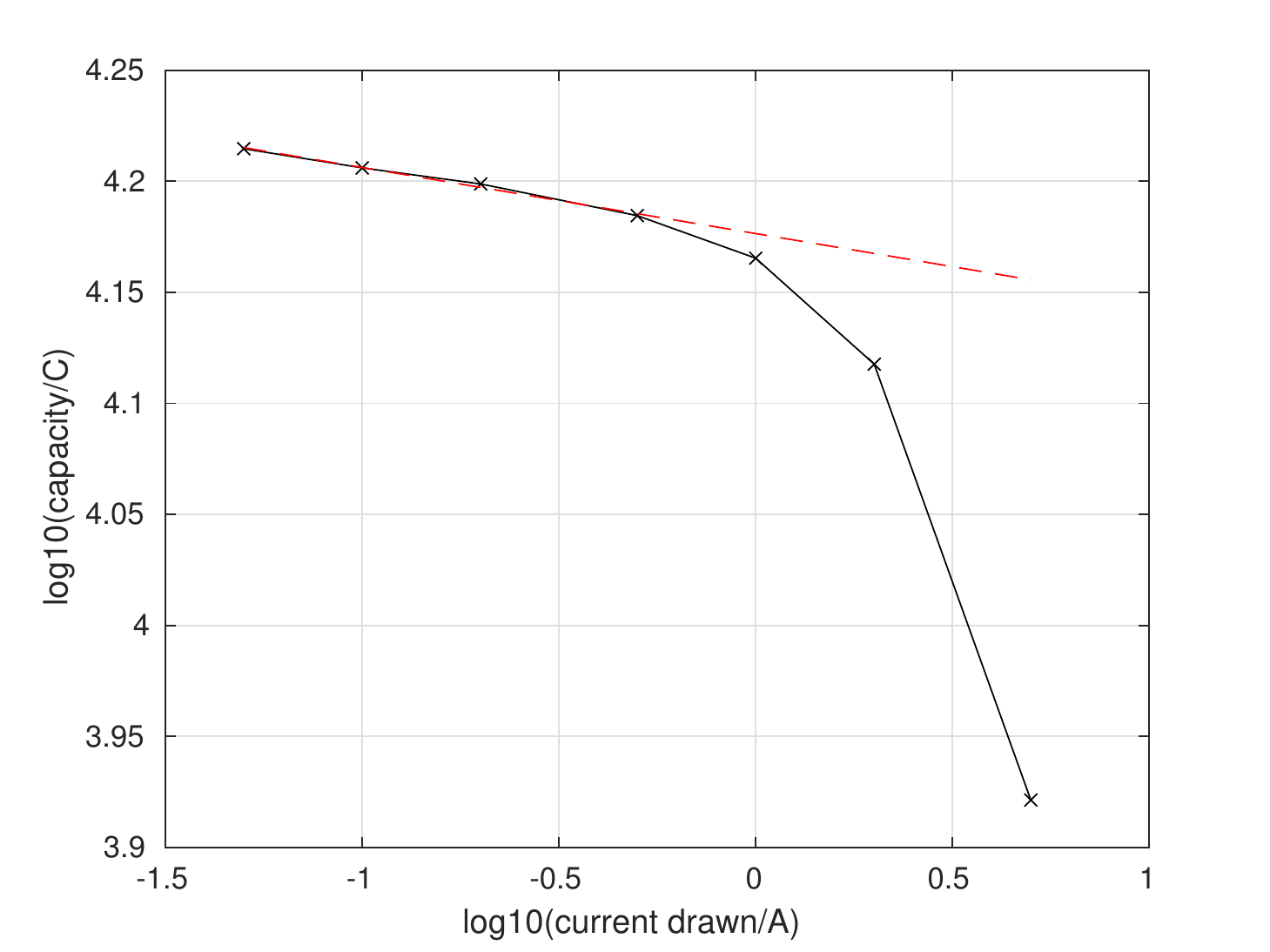}
 	\caption{The battery capacity against current drawn on a log-log plot (black line). The red dashed line, fitted to the lowest four discharge rate
 		datapoints, demonstrates the power-law behavior at 
 	low discharge rates.}
 	\label{fig:capacity_result}
 \end{figure}

Using the experimental values for $\alpha$, $C_F$ and $R_s$, we have simulated the voltage response of the battery
under the charge-discharge cycles at the chosen currents. The simulated voltage against time results are shown
in Fig.~\ref{fig:V_vs_t} as the dashed lines; the simulated capacity versus current results are shown in Fig.~\ref{fig:Q_vs_I},
again as a dashed line. Overall, the measured and modeled broadly agree, but there are some notable discrepancies. 
First, we note that the experimental voltage traces have a rapid drop in voltage as the voltage nears its lower
limit of 3.0~V. This is not replicated by the CPE-R model. As a result, the experimental plots feature more hysteresis.
Secondly, there are regular fluctuations about the voltage trend, for the experimental plots, particularly at the lower currents. These can be seen in Fig.~\ref{fig:V_vs_t}, near 4.3~V, as small amplitude waveform. 
It is possible that these fluctuations are a result of the regular layering of ions near the electrodes at the low currents~\cite{huggins}.
Finally, the characteristic `curve' of a CPE's voltage under a constant current~\cite{westerlund_dead_1991} is not evident because 
the exponent $\alpha$ is very close to 1 for this battery.

 \section{Discussion and Conclusions}
 
 We have used a constant current charge-discharge cycle to measure, at slow cycle rates, the capacity of an NCA battery at different currents. 
Figure~\ref{fig:Q_vs_I} demonstrates an increased capacity at the lowest currents. By modeling the battery as a CPE in series
with a resistor, we have been able to use fractional calculus as in Eq.~(\ref{eq:Qcalc}) to identify the exponent and fractional
capacity, $\alpha$ and $C_F$ respectively, and series resistance $R_s$. The modeled results using
a description of the CPE as a network of RC elements (dashed line in 
Fig.~\ref{fig:Q_vs_I}) agree well with the
measured results (solid line), although they are lower at the higher currents. 

Modeling of the voltage against time for the charge-discharge cycles, the dashed lines of Fig.~\ref{fig:V_vs_t}, shows general agreement with measured results (the solid lines in the same figure).
However, the experimental data shows more subtlety in variation with time. 
First, there is a rapid drop in voltage as the battery nears its lower limit $V_l$. This is a non-linear effect
that the linear CPE-R model is unable to reproduce. 
It might be possible to model this as an increase in effective
resistance at the lower voltages.
Secondly, the experimental data shows regular fluctuations during charge and discharge, especially at the lower currents,
as also seen in \cite{osara21}. 
These fluctuations or ``wiggles'' in the voltage plotted against time (or charge) are attributed to variations in the process of lithium intercalating into nanopores within disordered carbon of negative electrodes. The effect is expected to vary with the organisation and structure of the carbon, which itself depends upon the processing of the carbon in the manufacture of the battery. This explanation is consistent with the wiggles becoming more pronounced at lower current density. The interested reader is directed to chapter~7 of~\cite{huggins}.

The  values of $\alpha$ and $R_s$ found from the charge capacity, 0.9711(17) and 63~m$\Omega$ respectively, are in agreement with the values found from
the impedance plot of Fig.~\ref{fig:impedancef}, 0.976(8) and 57~m$\Omega$. We note
that the uncertainty in $\alpha$ is lower for the capacity method than for the impedance method, 
though we have only measured one chemistry and so we cannot generalize this result further at present.

However, the fractional capacity is lower than predicted from the impedance
measurements. This is a known phenomenon, especially visible in the case of lead-acid batteries~\cite{karden2011_ZleadAcid}. 

A notable feature absent in our modeling is the Warburg element. This is clearly evident on the impedance plot
of Fig.~\ref{fig:impedancef}(c), at high frequencies (right-hand half), as a line of angle approximately $-45^\circ$ 
to the horizontal axis. It is also evident in part (a) as a less sharp corner between the high-frequency and low-frequency
asymptotes that expected for a capacitor-resistor series pair of elements. 
However, we note that we have not required the Warburg to reproduce the up-tick of capacity at the lowest currents
in Fig.~\ref{fig:Q_vs_I}.

Finally, we note that this NCA battery has a power $\alpha$ very close to 1. This means it behaves almost as 
an ideal capacitor. While this is clearly an advantage for automotive applications, it means that demonstration of the CPE-nature
is harder to achieve than for `more fractional' batteries such as lead-acid. Yet the effect of the fractionality is 
still clearly evident through the uptick in the capacity at very low currents (Fig.~\ref{fig:Q_vs_I}). In summary,
the application of long timescale measurements to a battery are able to elucidate its fractional behavior.

 \section*{Appendix}
 
 In principle, a CPE can be modeled using an infinite array of RC elements in parallel, as first introduced by Morrison~\cite{morrison_rc_1959}. Resistor and capacitor values can be chosen so that the array behaves as a CPE with a chosen exponent $\alpha$ and fractional capacitance $C_F$. The underlying principle is that the RC time constants of the branches increase
 logarithmically from one branch to the next, meaning that at any frequency only a few branches of the array will have appreciable admittance; appropriately chosen  R and C values ensure the desired power-law response for admittance (or impedance) and constant phase with changing frequency. 
 
 In practice, of course, any such modeled network needs to be finite. Specifically, if we define the resistance of the central `zero-th' branch as $R_0$, and the capacitance as $C_0$,
 the resistance of the $i$-th branch ($i=-N \cdots N$) 
 is $R_i = R_0 k^i$ and the capacitance of the $i$-th branch is
 $C_i = C_0 [k^{1/\alpha - 1}]^i$, where $k$ is a multiplier given by $k=k_f^\alpha$ with $k_f > 1$.  Here $k_f$ describes the resolution of the representation; as
 $k_f \rightarrow 1$ and $N \rightarrow \infty$ the model
 gives a pure CPE behavior. In practice, however, a discrete network is adequate. 
 
 A complication of this approach is that the minimum time step required to simulate the network  is defined by the smallest RC  time constant, and thus $N$ is practically limited by available CPU time. To avoid requiring an unfeasibly large
 $N$, Morrison showed how the effect of the branches with
 smaller time constants can be approximated by a single capacitor in parallel with the network --- in effect making the assumption that the 
 capacitors in the low time constant branches respond
 instantly to a change in voltage. This extra capacitor
 $C_t$ has value:
 \begin{equation}
 C_t = C_{-N} \frac{ k^{1/\alpha-1} }{ k^{1/\alpha-1}-1}.
\end{equation} 
 
\begin{figure}[tbp]
	\centering
	\includegraphics[width=0.95\linewidth]{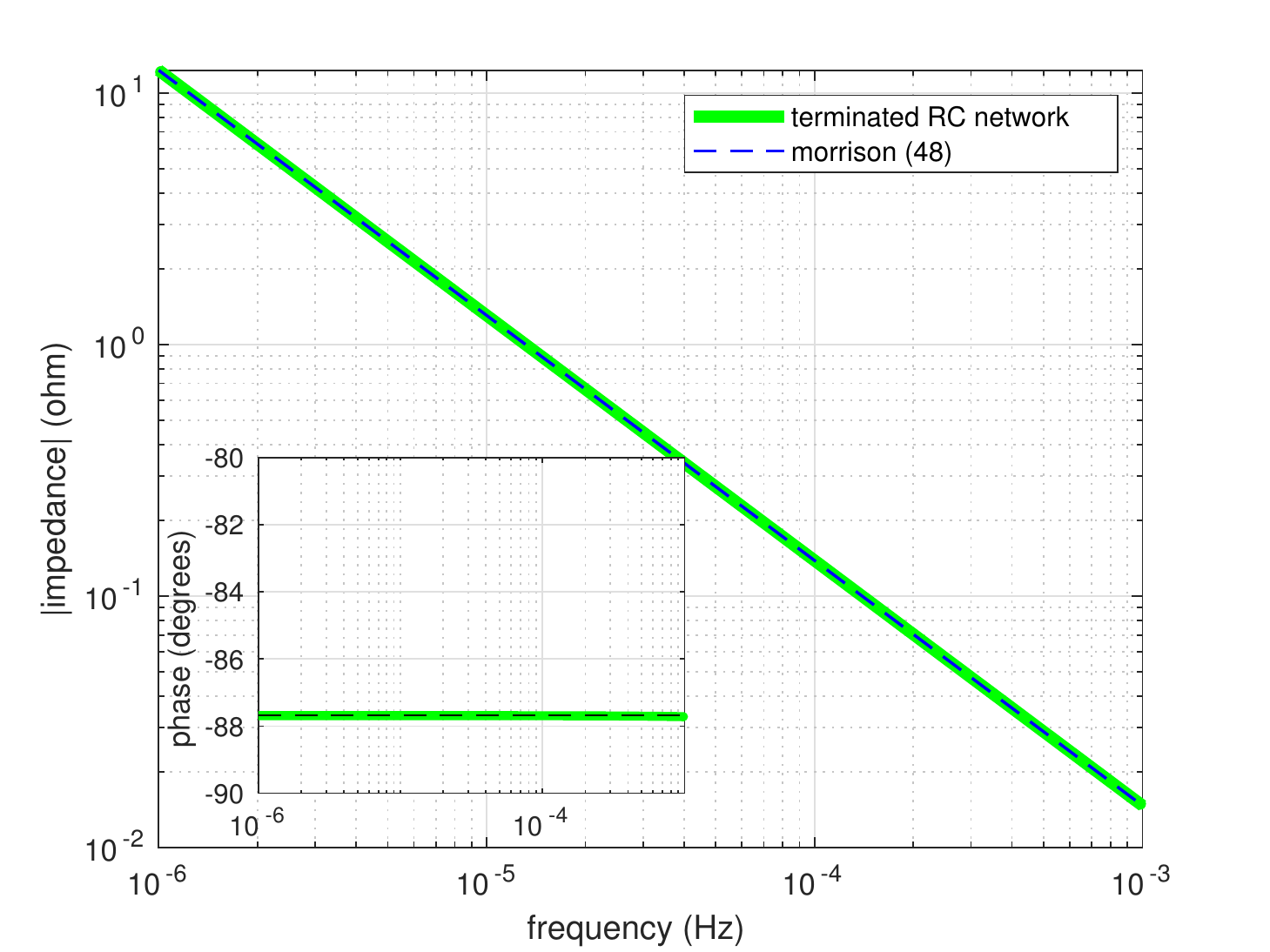}
	\caption{A comparison of the modeled impedance with a Morrison network (green solid line) and the required CPE impedance (black dash line).}
	\label{fig:morrison_comp}
\end{figure}

 For the modeling in this paper, we have used
 a network with $N=30$ (i.e. 61 branches), $k_f=1.4$ with
 a center branch having $R_0 = 725~\Omega$ and $C_0 = 110$~F. The terminating capacitor has $C_t = 9840 F$. This gives the required CPE of $\alpha = 0.9711$ and $C_F = 9203$~A~s$^{\alpha}$~V$^{-1}$. A plot of the
 impedance of the network, along with the desired impedance from
 Eq.~(\ref{eq:Zdef}) is shown in Fig.~\ref{fig:morrison_comp}. The model reproduces the required impedance
 very accurately over the full range of time-scales used in our experimental measurements.
 The current in
 each branch, and the voltages over the capacitors 
 against time are solved using Matlab and iterated forward in time in
 steps of 1~s, about 0.25 times the lowest branch RC time constant. 
 
\bibliographystyle{elsarticle-num.bst} 
\bibliography{BatteryLibrary.bib}

\end{document}